\documentclass[12pt,a4paper,fleqn]{article}
\usepackage{umlaut,graphicx,epsfig,epsf,rotating}

\newcommand{\etal}{{\sl et. al.}}
\newcommand{\beq}{\begin{equation}}
\newcommand{\eeq}{\end{equation}}

\begin{document}
\thispagestyle{empty}

\noindent
 {\bf Corresponding author}\\
Prof. Dr. H.V. Klapdor-Kleingrothaus\\
Max-Planck-Institut f\"ur Kernphysik\\
Saupfercheckweg 1\\
D-69117 HEIDELBERG\\
GERMANY\\
Phone Office: +49-(0)6221-516-262\\
Fax: +49-(0)6221-516-540\\
email: $klapdor@gustav.mpi-hd.mpg.de$\\

\begin{center}
{\Large \bf 
Background Analysis around Q$_{\beta\beta}$ for $^{76}{Ge}$ 
	Double Beta Decay experiments, and Statistics 
	at Low Count Rates}\\
\vspace{1.5cm}
{ H.V. Klapdor-Kleingrothaus${}^1$, A. Dietz${}^1$,
I.V. Krivosheina${}^{1,2}$, C.~D\"orr${}^1$, C.~Tomei${}^{3}$}\\
{
\vspace{0.75cm}
{\sl ${}^1$Max-Planck-Institut f\"ur Kernphysik, P.O. 10 39 80}\\
{\sl D-69029 Heidelberg, Germany} \\
{\sl ${}^2$Radiophysical-Research Institute, Nishnii-Novgorod,
Russia}\\
{\sl ${}^3$Universita degli studi di L'Aquila, Italy}
}
\end{center}

\vspace{1.5cm}

\begin{abstract}

The background in the region of the Q-value for neutrinoless 
	double beta decay of $^{76}{Ge}$  has been investigated 
	by different methods: Simulation with GEANT 4 
	of the HEIDELBERG-MOSCOW experiment, analysis of other Ge 
	double beta experiments. Statistical features of the analysis 
	at very low count rates are discussed. 

\end{abstract}

\newpage

\section{Introduction}

	Recently first experimental evidence has been reported for neutrinoless
	double beta decay. Analysis of 55 kg y of data, taken by the
	HEIDELBERG-MOSCOW experiment in the GRAN SASSO 
	over the years 1990 - 2000,
	has led \cite{evid1,evid2,evid3,evid4} to a half-life

\beq
T_{1/2} = (0.8-18.3)\times10^{25} \quad\mbox{years}\quad  (95\% \;C.L.),
\eeq

	with best value of $T_{1/2}=1.5\times 10^{25}$ y,
	for the decay of the double beta emitter $^{76}$Ge

\beq
^{76}\mbox{Ge} \longrightarrow ^{76}\mbox{Se} +2e^-
\eeq

	Assuming the decay amplitude to be dominated by exchange of a massive
	Majorana neutrino (see, e.g. \cite{60years}), 
	this half-life results in a value
	of the effective neutrino mass

\beq
<m> = \left|  \sum U^2_{ei} m_i\right| = 0.05 - 0.84\; \mbox{eV} \quad(95\% \;C.L.),
\eeq

	with best value of 0.39 eV.
	Here a 50\% uncertainty in the nuclear matrix elements 
	has been taken into
	account (for details see \cite{evid3}).\\

\begin{figure}[thb]
\begin{center}
\begin{sideways}
\begin{sideways}
\begin{sideways}
\includegraphics[width=7.0cm]{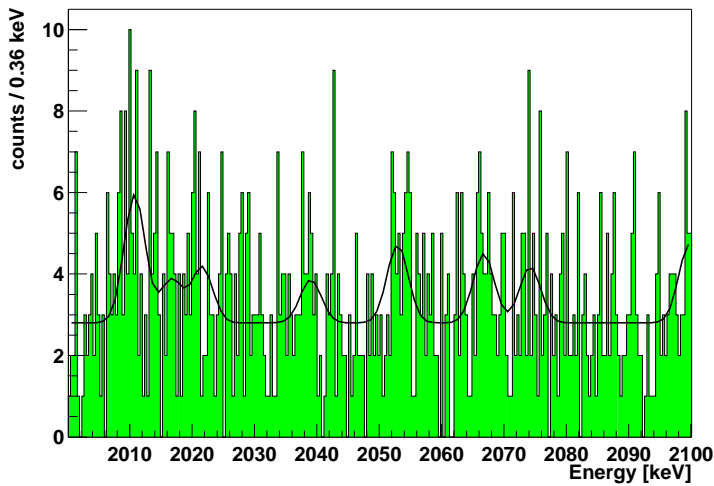}
\end{sideways}
\end{sideways}
\end{sideways}
\end{center}
\caption{\label{picMany}\rm \small The spectrum taken with $^{76}$Ge detectors 
	Nrs. 1,2,3,4,5 over the period August 1990 - May 2000 
	(54.98 kg y), in the energy range 2000 - 2100 keV. 
	Simultaneous fit of the $^{214}$Bi-lines and the two high-energy
	lines yield a probability for a line at 2039 keV of 91\% C.L.}
\end{figure}
\begin{figure}[th]


\begin{center}
\includegraphics[width=6cm]{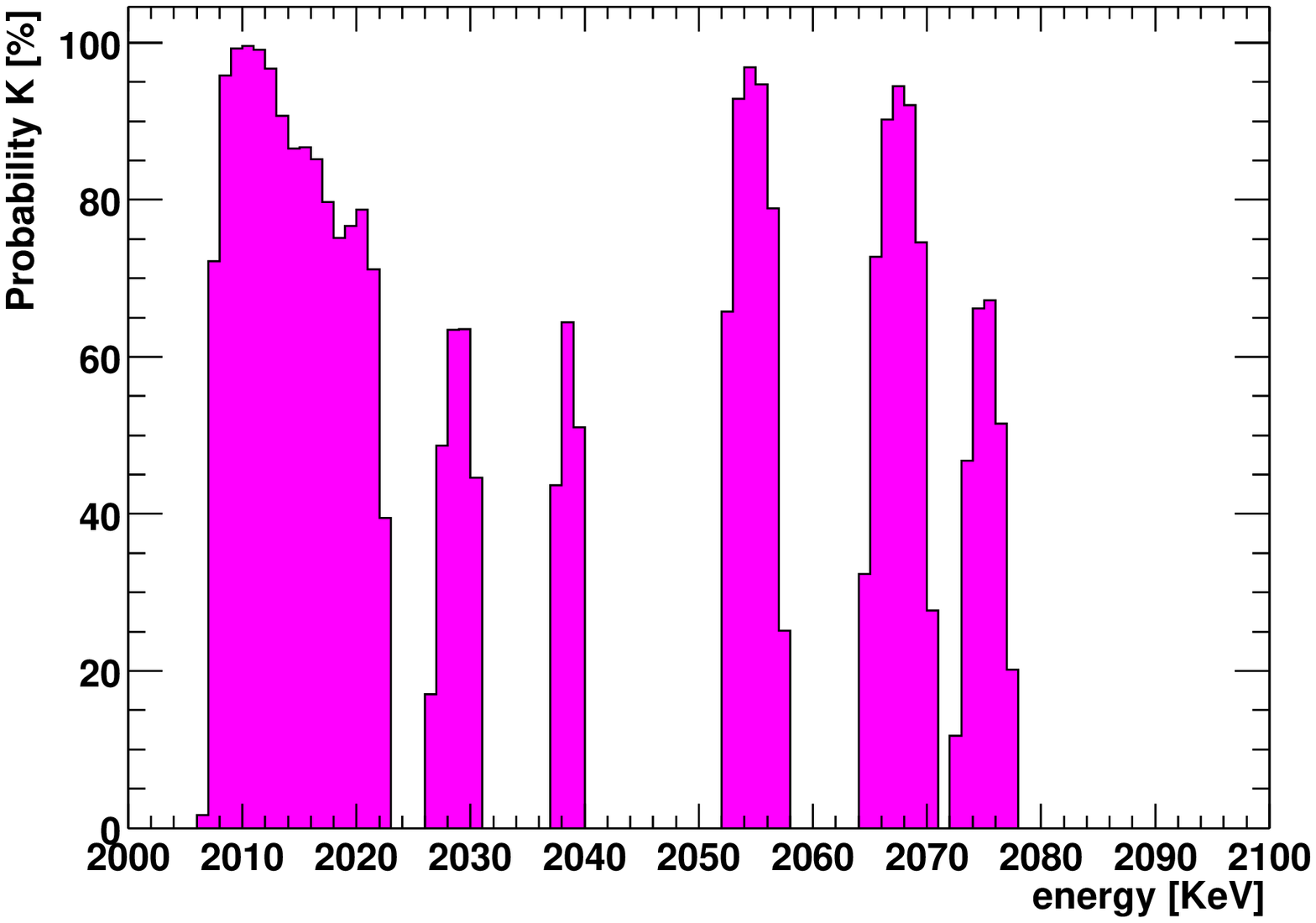}
\vspace*{0cm}
\includegraphics[width=6cm,height=5cm]{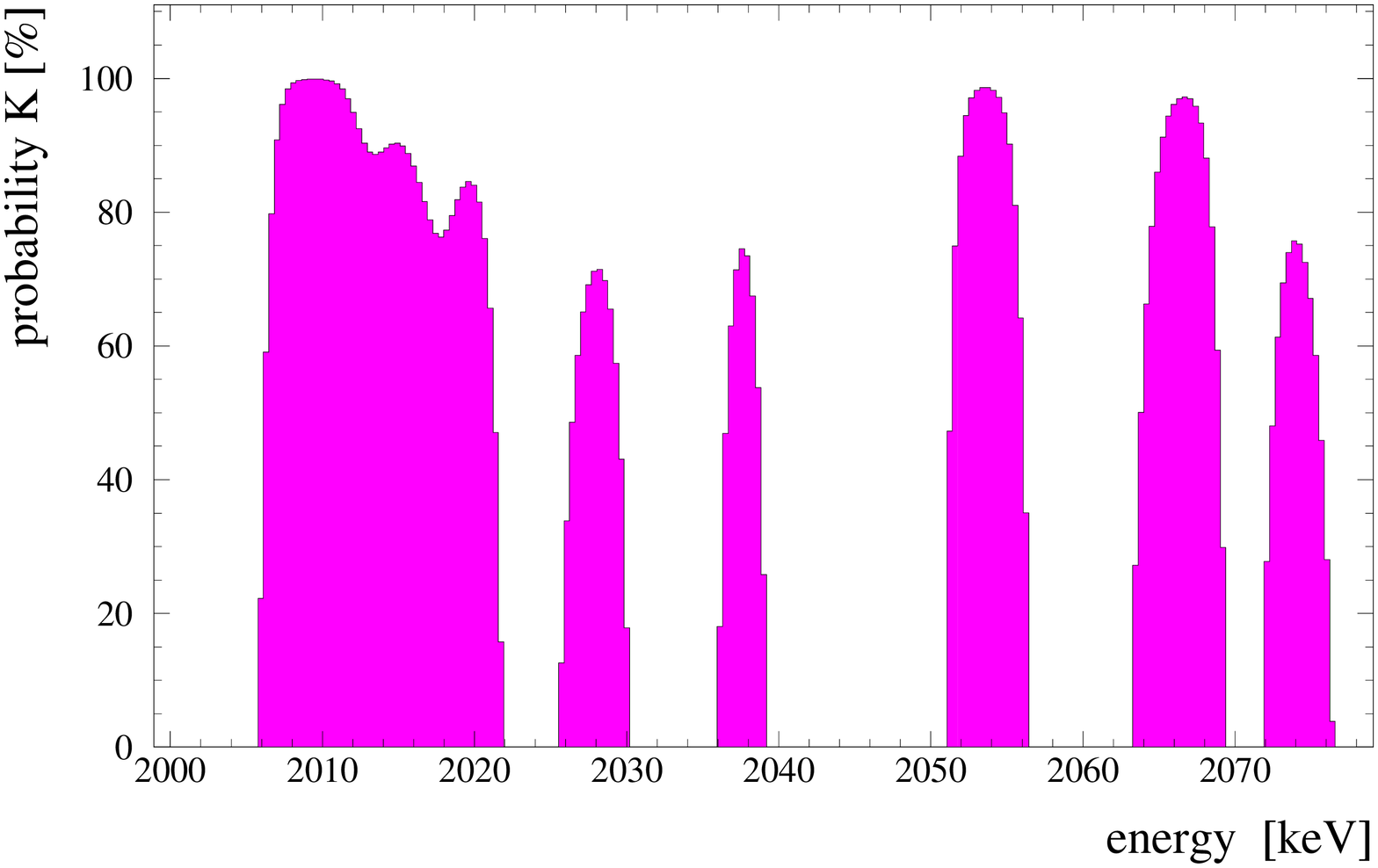}
\caption{\label{figHM}\rm \small Result of the peak-searching procedure
	performed on the HEIDELBERG-MOSCOW spectrum using the Maximum
	Likelihood approach (left) and the Bayesian method (right).
	On the y axis the probability 
	of having a line at the corresponding energy in the spectrum is shown.}
\end{center}
\end{figure}

\begin{figure}[h!]
\begin{center}
\includegraphics[width=6.0cm]{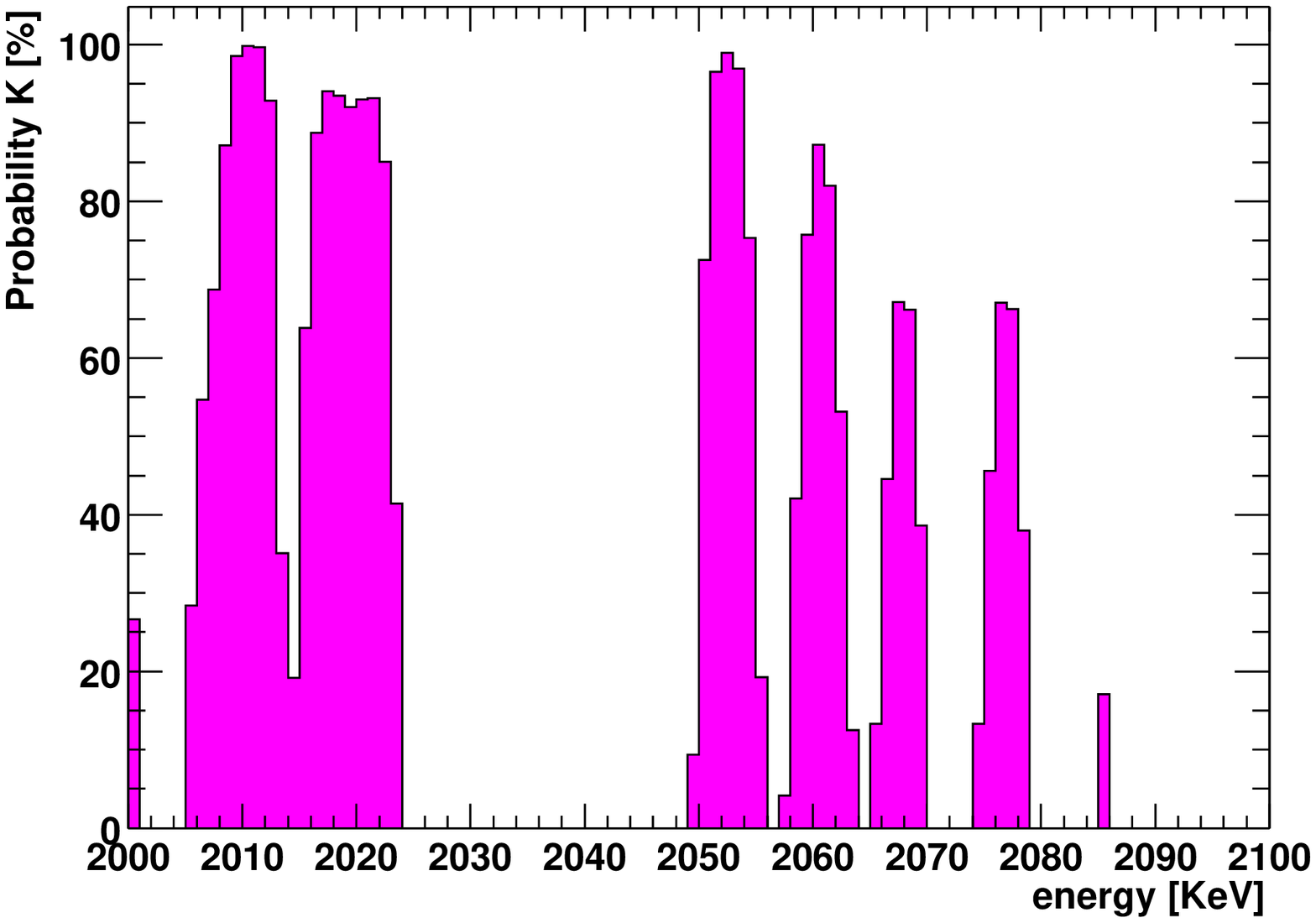}
\includegraphics[width=6.0cm]{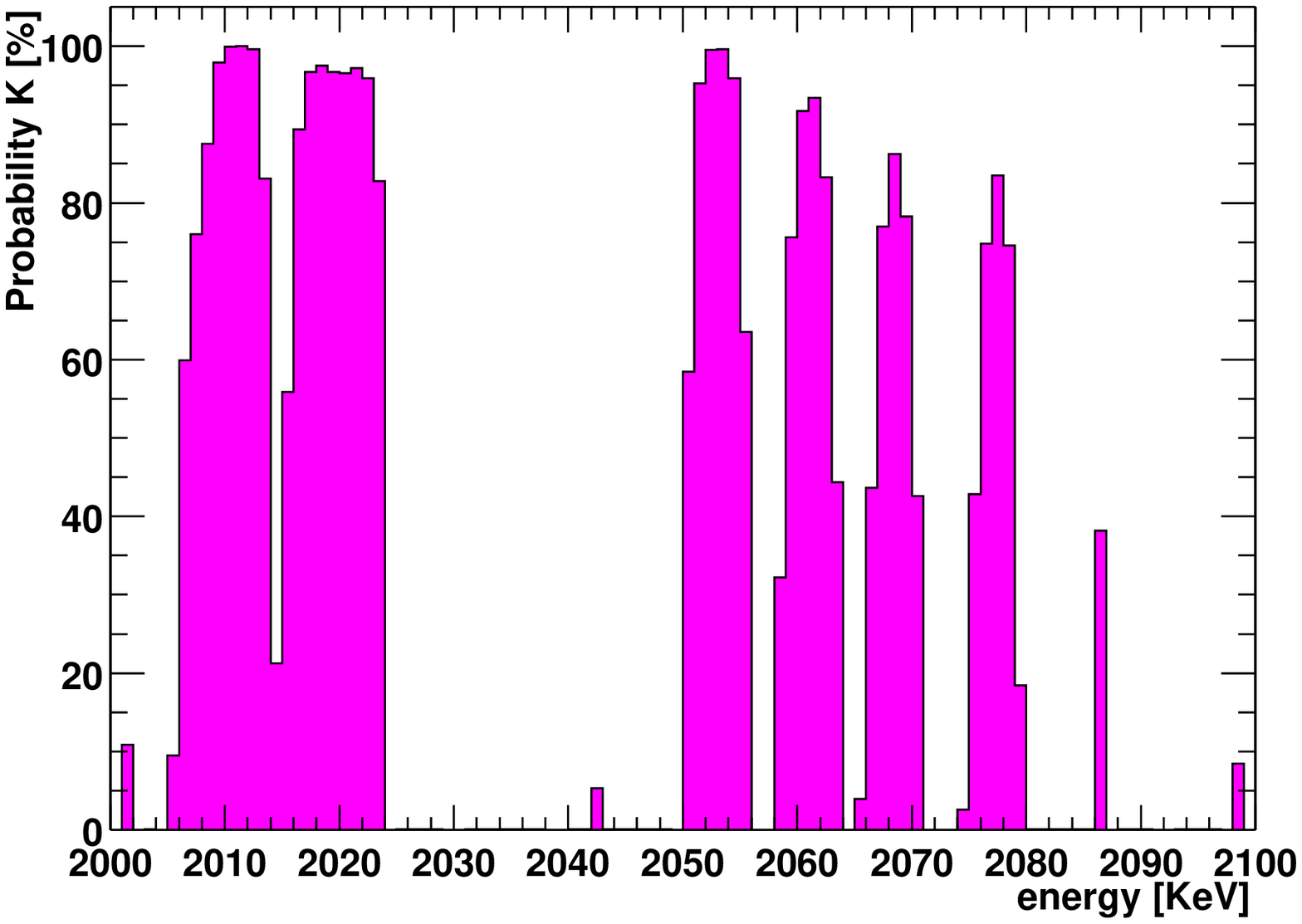}
\caption{\label{figCaldwell}\rm \small Result of the peak-search procedure
	performed for the UCBS/LBL spectrum \cite{caldwell} (left: Maximum
	Likelihood method, right: Bayes method). On the y axis the probability 
	of having a line at the corresponding energy in the spectrum is shown.}
\end{center}
\end{figure}


	This is for the first time that the absolute scale of the neutrino mass
	spectrum has been fixed, which cannot be achieved by neutrino 
	oscillation experiments. This result restricts possible neutrino 
	mass scenarios to
	degenerate or (still marginally allowed) inverse hierarchy
\cite{con1,con2,con3}. 
	In the degenerate case it leads to a common
	neutrino mass eigenvalue of

\beq
m_1 = 0.05 - 3.4\; \mbox{eV}  \quad(95\%\; C.L.).
\label{eq1}
\eeq

	This result is nicely consistent with later collected or analyzed
	experimental data, such as results from Large Scale Structure
	and CMB measurements 
\cite{cmb1,cmb2,cmb3}, 
	or ultra-high energy cosmic rays 
\cite{cray}.
	The former yield an upper limit of $\sum_i m_i$=1.0 eV 
	(corresponding in the degenerate case to a common mass 
	eigenvalue $m_0 < 0.33$ eV).
	The Z-burst scenario for ultra-high energy cosmic 
	rays requires 0.1 - 1.3 eV  
\cite{cray}. 
	Tritium single beta decay cuts the upper range in
	eq.(\ref{eq1})  down to 2.2 or 2.8~eV 
\cite{tritium}.
\\

	There is further theoretical support for a neutrino mass in the range
	fixed by the HEIDELBERG-MOSCOW experiment. A model based 
	on an A4 symmetry of the neutrino mass matrix requires the
	neutrinos to be degenerate and the common mass eigenvalue to be
	$>$0.2~eV 
\cite{ma}. 
\\

	Starting with the hypothesis that quark and lepton mixing are
	identical at or near the GUT scale, Mohapatra \etal 
~\cite{mohap} show
	that the large solar and atmospheric neutrino mixing angles can be
	understood purely as result of renormalization group evolution, if
	neutrino masses are quasi-degenerate (with same CP parity). The common 
	Majorana neutrino mass then must be larger than 0.1 eV.\\

\begin{figure}[t]
\begin{center}
\includegraphics[width=6cm]{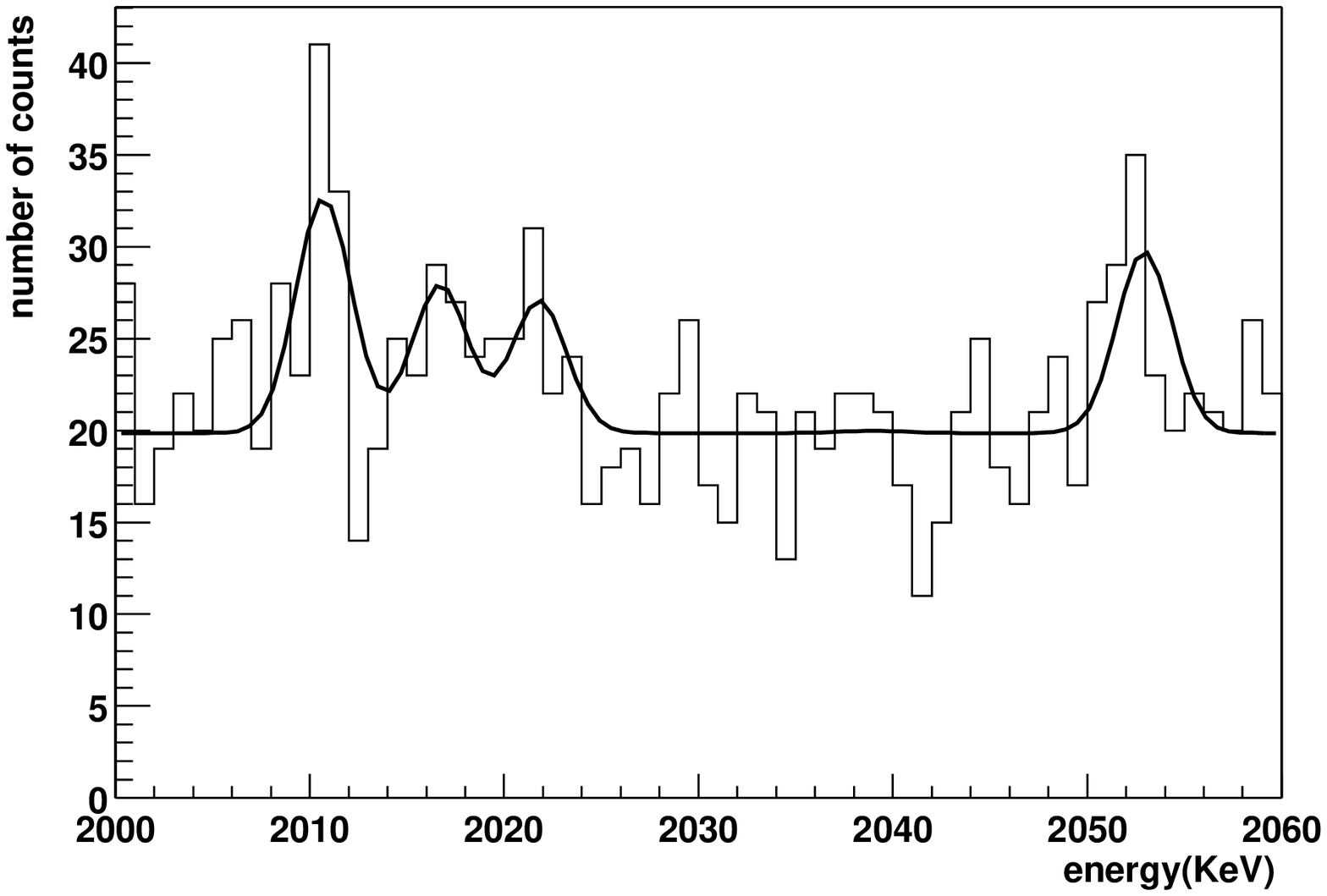}
\caption{\label{specCaldwell}\rm \small 
	Analysis of the spectrum measured by D. Caldwell et al. 
\cite{caldwell},
        with the Maximum Likelihood Method, in the energy 
        range 2000 - 2060\,keV 
        assuming lines at 2010.78, 2016.70, 2021.60, 2052.94,
	2039.0\,keV.        
        No indication for a signal at 2039\,kev is observed in this case.}
\end{center}
\end{figure}

	In this paper we report additional support of the double beta result of
\cite{evid1,evid2,evid3,evid4}, 
	by further discussion of the structure 
	of the experimental background, by statistical considerations 
	and by analysis of other double
	beta experiments investigating the decay of $^{76}$Ge.\\

	Important points in the analysis of the measured spectrum are the
	structure of the background around $Q_{\beta\beta}$ (=2039.006(50) keV 
\cite{qvalue}), 
	and the
	energy range of analysis around $Q_{\beta\beta}$.

\section{Structure of Background Around Q$_{\beta\beta}$ 
in Different Ge Double Beta Experiments}

	Fig. 1 shows the spectrum measured in the range 2000 - 2100 keV in its
	original binning of 0.36 keV. By the peak search procedure 
	developped 
\cite{evid2,evid3}
	on basis of the Bayes and Maximum Likelihood Methods, exploiting as
	important input parameters the experimental knowledge on the shape and
	width of lines in the spectrum, weak lines of $^{214}$Bi have 
	been identified at the energies of 2010.78, 2016.7, 2021.6 
	and 2052.94 keV 
\cite{evid1,evid2,evid3,evid4}.
	Fig. 2 shows the probability that there is a line of correct 
	width and of Gaussian shape at a given energy, assuming all 
	the rest of the spectrum as
	flat background (which is a highly conservative assumption), 
	for the HEIDELBERG-MOSCOW experiment.\\

	The intensities of these lines have been shown to be consistent 
	with other, strong Bi lines in the measured spectrum according 
	to the branching ratios
	given in the Table of Isotopes 
\cite{toi}, 
	and to Monte Carlo simulation of the experimental setup 
\cite{evid3}. 
	Note that the 2016 keV line, as an E0 transition,
	can be seen only by coincident summing of the two successive lines
	$E=1407.98$ keV
	and $E=609.316$ keV. Its observation proves that the $^{238}$U 
	impurity from which
	it is originating, is located in the Cu cup of the detectors.
	Recent measurements of the spectrum of a $^{214}$Bi {\it source } as
	function of distance source-detector {\it confirm} 
	this interpretation 
\cite{oleg}.
\\

	Premature estimates of the Bi intensities given in Aalseth et.al,
	hep-ex/0202018 and Feruglio et.al., Nucl. Phys. {B 637} (2002), 345,
	are incorrect, because this long-known spectroscopic
	effect of true coincident summing 
\cite{gamma} 
	has not been taken into account, and also no simulation 
	of the setup has been performed (for details see 
\cite{evid3,kk0205}). 
\\

	These $^{214}$Bi lines occur also in other investigations 
	of double beta decay
	with Ge - and - even more important - also the additional structures in
	Fig. 2, which cannot be attributed  at present, are seen in these
	other investigations.\\

	There are three other Ge experiments which have looked for double beta
	decay of $^{76}$Ge. First there is the experiment by Caldwell et al. 
\cite{caldwell}, 
	using natural Germanium detectors (7.8\% abundance of $^{76}$Ge, 
	compared to 86\% in
	the HEIDELBERG-MOSCOW experiment). 
	This was the most sensitive {\it natural} Ge
	experiment. With their background a factor of 9 higher than in the
	HEIDELBERG-MOSCOW experiment and their measuring time 
	of 22.6 kg y ears,
	they had a statistics of the background by a factor of almost four
	\mbox{l a r g e r} than in the HEIDELBERG-MOSCOW experiment. 
	This gives useful
	information on the composition of the background.

\begin{figure}[th]
\begin{center}
\includegraphics[width=6.0cm]{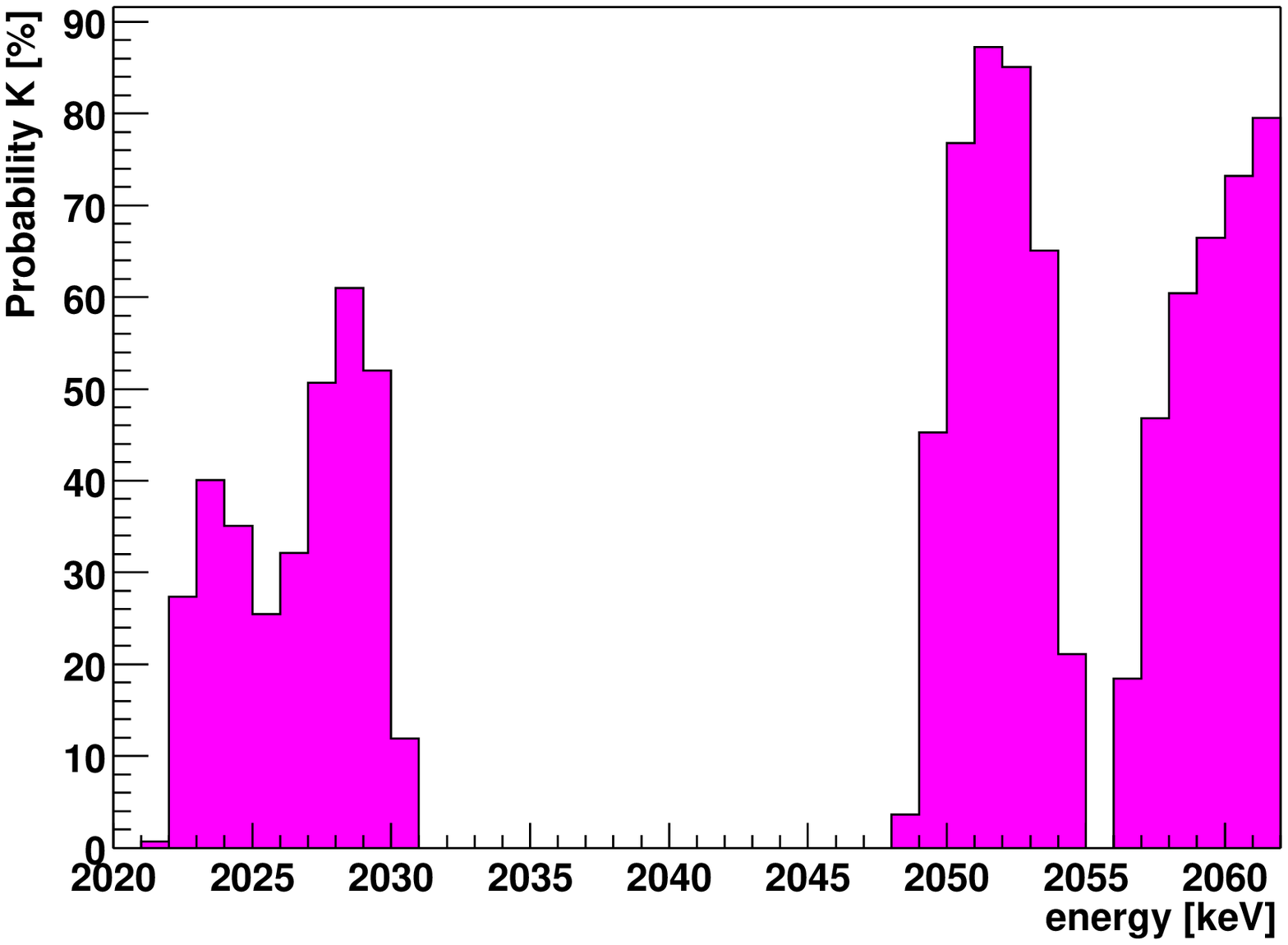}
\includegraphics[width=6.0cm]{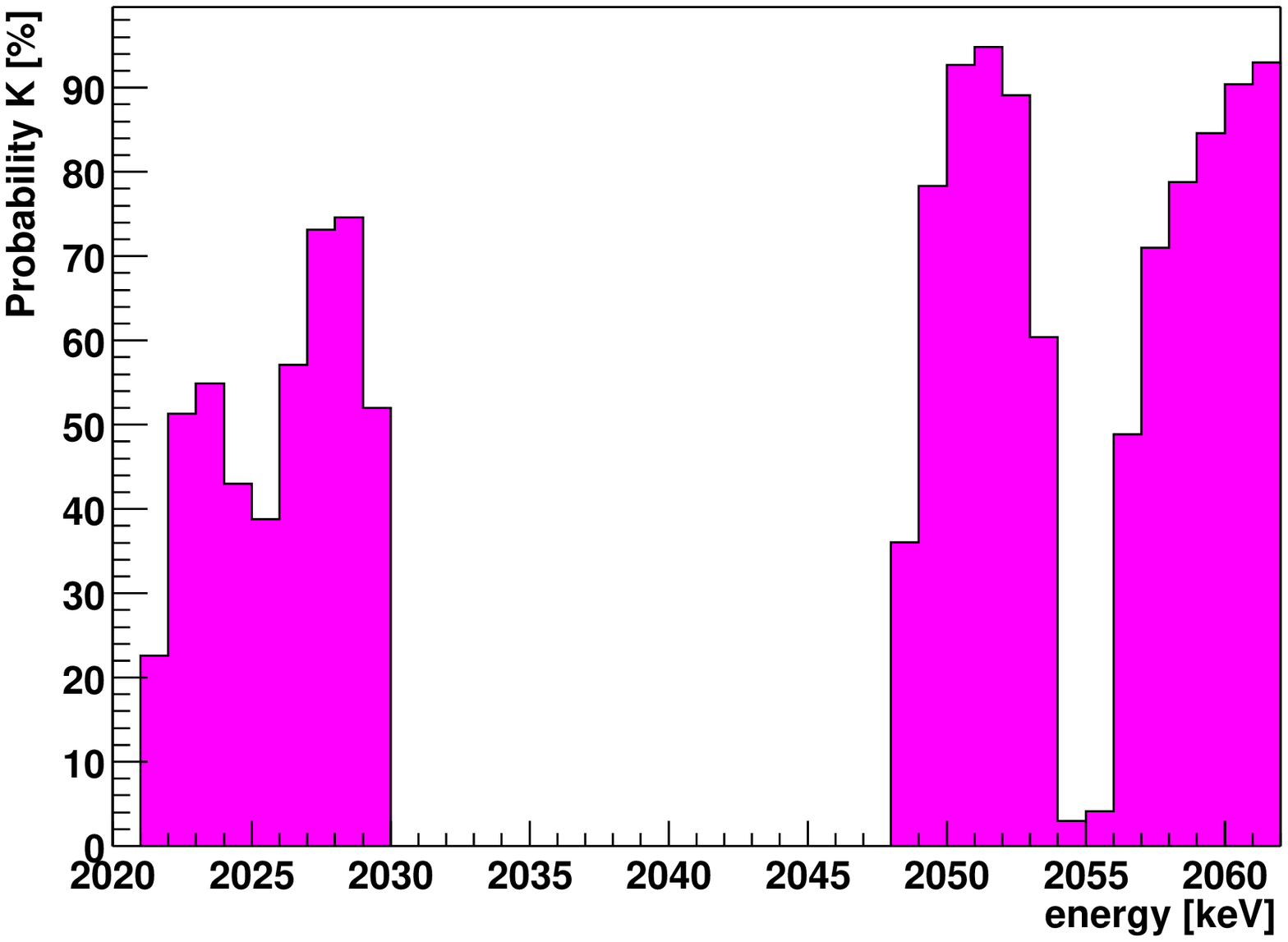}
\caption{\label{figITEP}\rm \small Result of the peak-search procedure
performed for the ITEP/YePI spectrum \cite{vasenko} (left: Maximum
Likelihood method, right: Bayes method). On the y axis the probability 
of having a line at the corresponding energy in the specrtum is shown.}
\end{center}
\end{figure}

\begin{figure}[h!]
\begin{center}
\includegraphics[width=6.0cm]{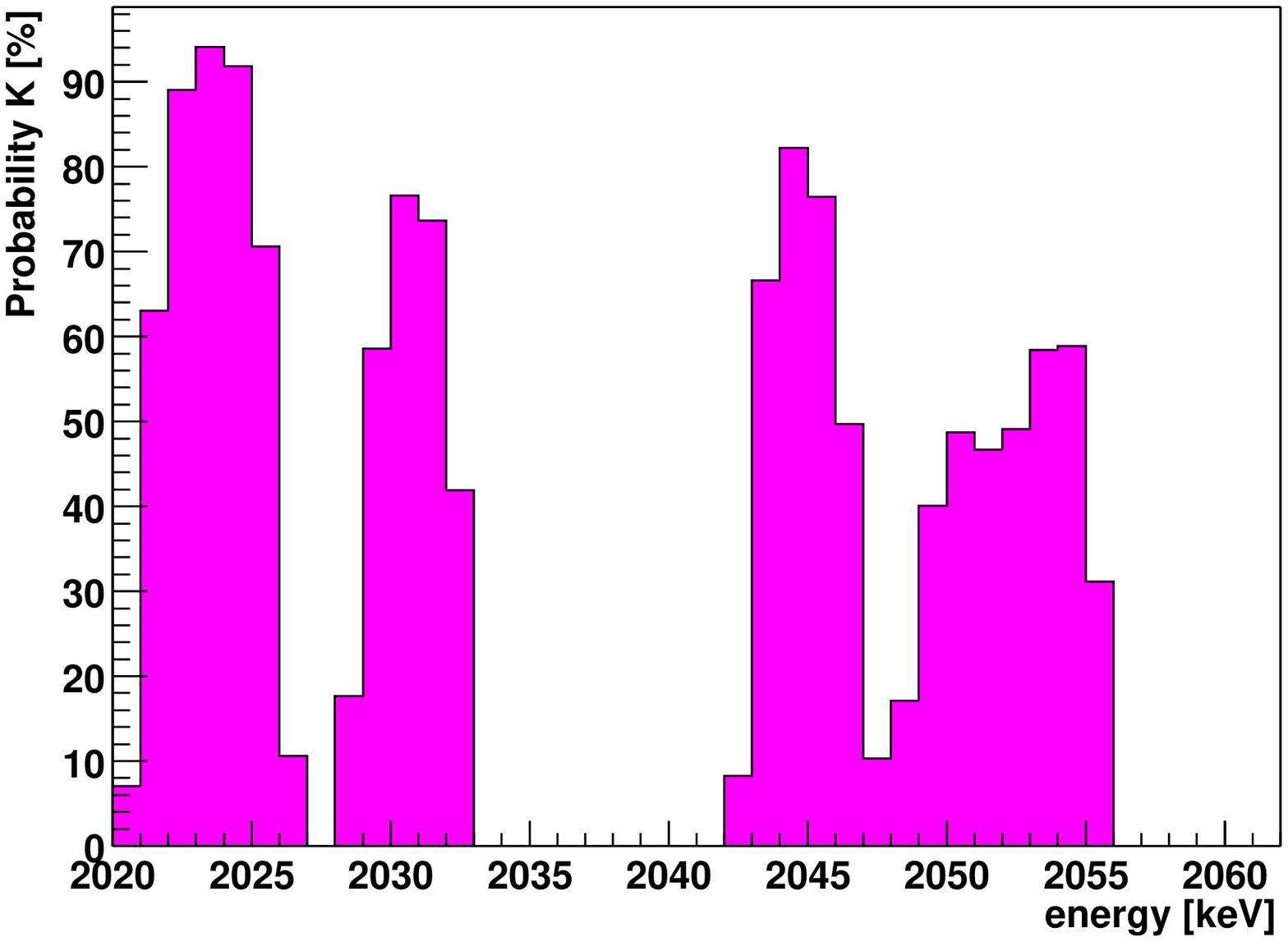}
\includegraphics[width=6.0cm]{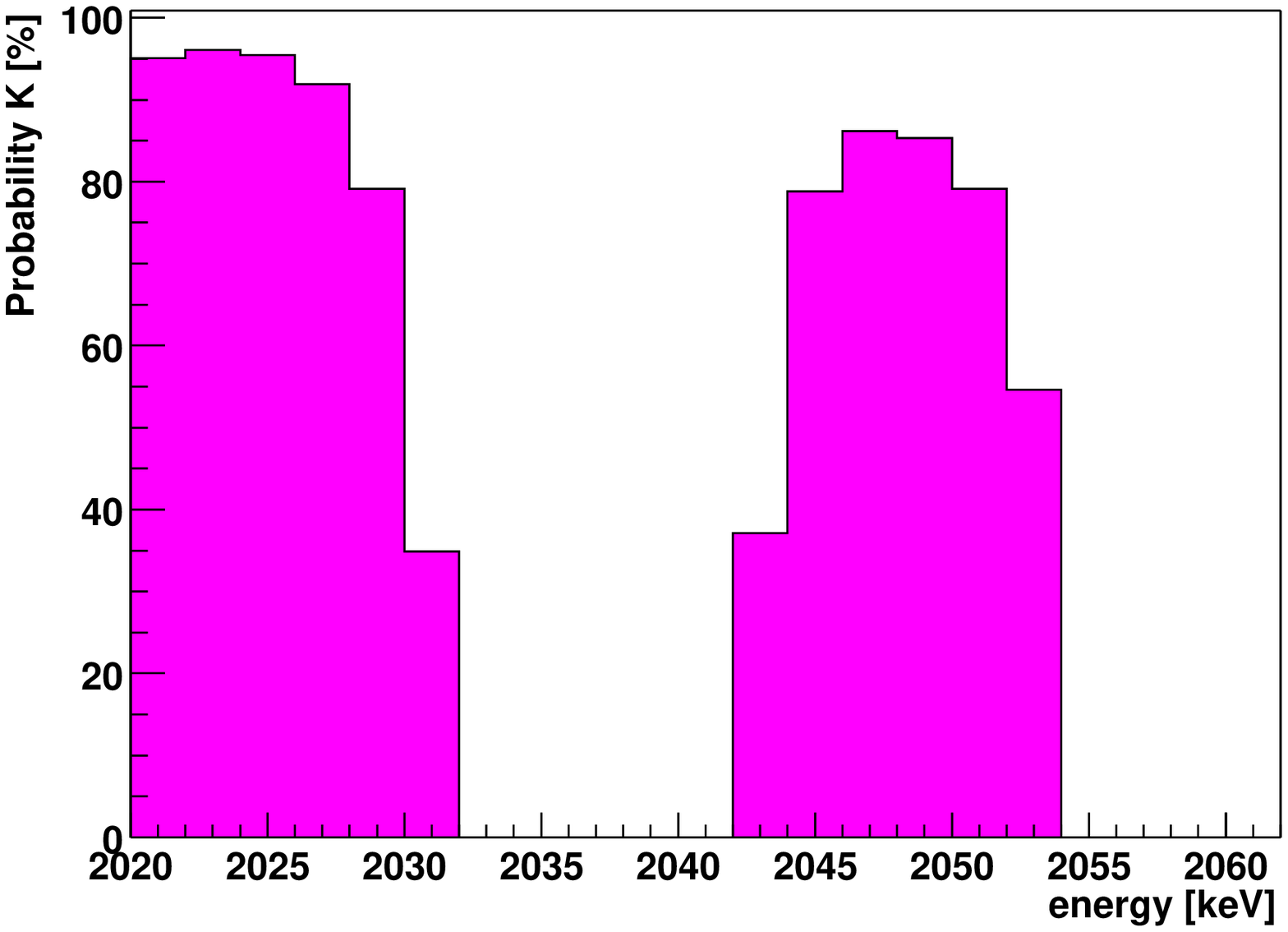}
\caption{\label{figIGEX}\rm \small Result of the peak-search procedure
performed for the IGEX spectrum \cite{igex,igex2} using the ML approach
(left) and the Bayesian statistics (right). On the y axis the probability 
of having a line at the corresponding energy in the spectrum is shown.}
\end{center}
\end{figure}


	Applying the same method of peak search as used in Fig. 
\ref{figHM}, 
	yields indications for peaks essentially at the same energies as in
	Fig. 
\ref{figHM} (see Fig. \ref{figCaldwell}).  
	This shows that these peaks are not fluctuations. In particular
	it sees the 2010.78, 2016.7, 2021.6 and 2052.94 keV $^{214}$Bi lines, 
	but a l s o  the unattributed lines at higher energies.  
	It finds, however, n o  line at 2039 keV.  This is consistent with the
	expectation from the rate found in the HEIDELBERG-MOSCOW experiment.
	About 16 observed events in the latter correspond to to 0.6  expected
	events in the Caldwell experiment, because of the use of non-enriched
	material and the shorter measuring time. Fit of the Caldwell spectrum
	allowing for the $^{214}$Bi lines and a 2039 keV line 
	yields 0.4 events for the latter (see Fig. 
\ref{specCaldwell}).\\

	The first experiment using enriched (but not high-purity) Germanium 76
	detectors was that of Kirpichnikov and coworkers 
\cite{vasenko}. 
	These authors show only the energy range between 2020 and 2064 keV of
	their measured spectrum.
	The peak search procedure finds also here indications of lines
	around 2028 keV and 2052 keV (see fig. 
\ref{figITEP}, 
	but \mbox{n o t} any indication of a line at 2039 keV. 
	This is consistent with the expectation, because for their low
	statistics of 2.95 kg y they would expect here (according to
	HEIDELBERG-MOSCOW) 0.9 counts.\\

	Another experiment (IGEX) used between 6 and 8.8 kg 
	of enriched $^{76}$Ge, but collected since beginning 
	of the experiment in the early nineties
	till shutdown in 1999 only 8.8 kg\,years of statistics 
\cite{igex,igex2}.
	The authors of 
	\cite{igex,igex2} 
	unfortunately show only the range 2020 to 2060 keV 
	of their measured spectrum in detail. Fig. 
\ref{figIGEX}  
	shows the result of our peak scanning of this range. 
	Clear indications are seen for
	the Bi lines at 2021 and 2052 keV, but also 
	of the unidentified structure around 2030 keV. 
	Because of the conservative assumption on the background
	treatment in the scanning procedure (see above) there 
	is no chance to see a signal at 2039 keV because of the 'hole' 
	in the background of that spectrum (see Fig. 1 in 
\cite{igex}). 
	With some good will one might see, however,
	an indication of 3 events here, consistent with the expectation of the
	HEIDELBERG-MOSCOW experiment of 2.6 counts.\\

	It may be noted that all three experiments with {\it enriched}
	$^{76}$Ge see a line around 2028 keV. 
	It was already mentioned in 
\cite{beyond}, 
	that it is suspicious that this line which is seen also 
	clearly in the pulse-shape
	analyzed spectrum of the HEIDELBERG-MOSCOW experiment 
	(see 
\cite{evid1,evid2,evid3})
	differs in energy from $Q_{\beta\beta}$ just by the K-shell 
	X-ray energies of Ge (Se)
	of 9.2 (10.50) keV, or the K shell electron atomic binding energies
	of 11.10  (12.66) keV (see 
\cite{toi}), 
	and that this might point to a partly not understood
	mechanism  of the double beta decay process.\\

\begin{figure}[t!]
\begin{center}
\includegraphics[width=6.7cm]{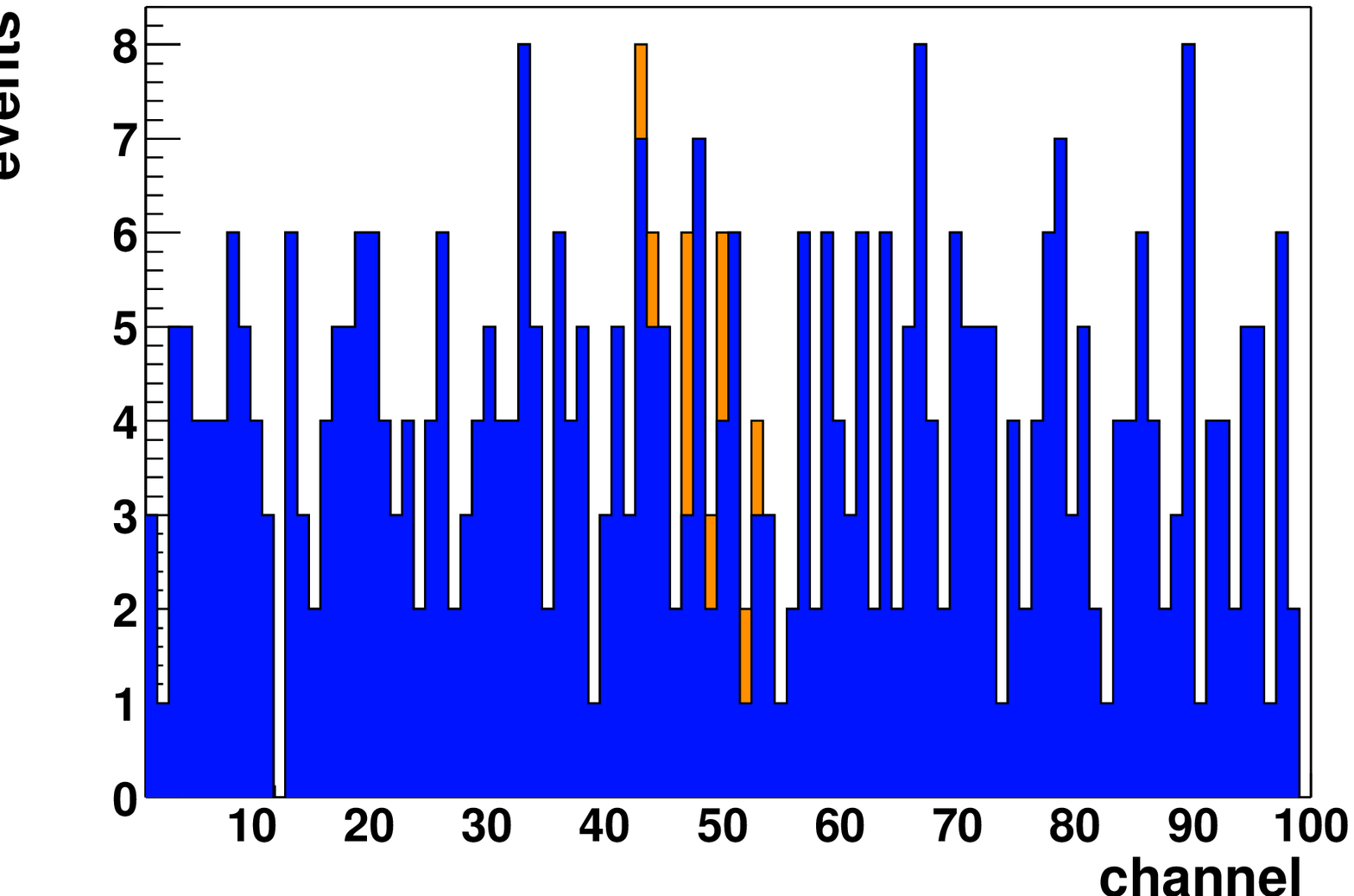}
\includegraphics[width=6.7cm]{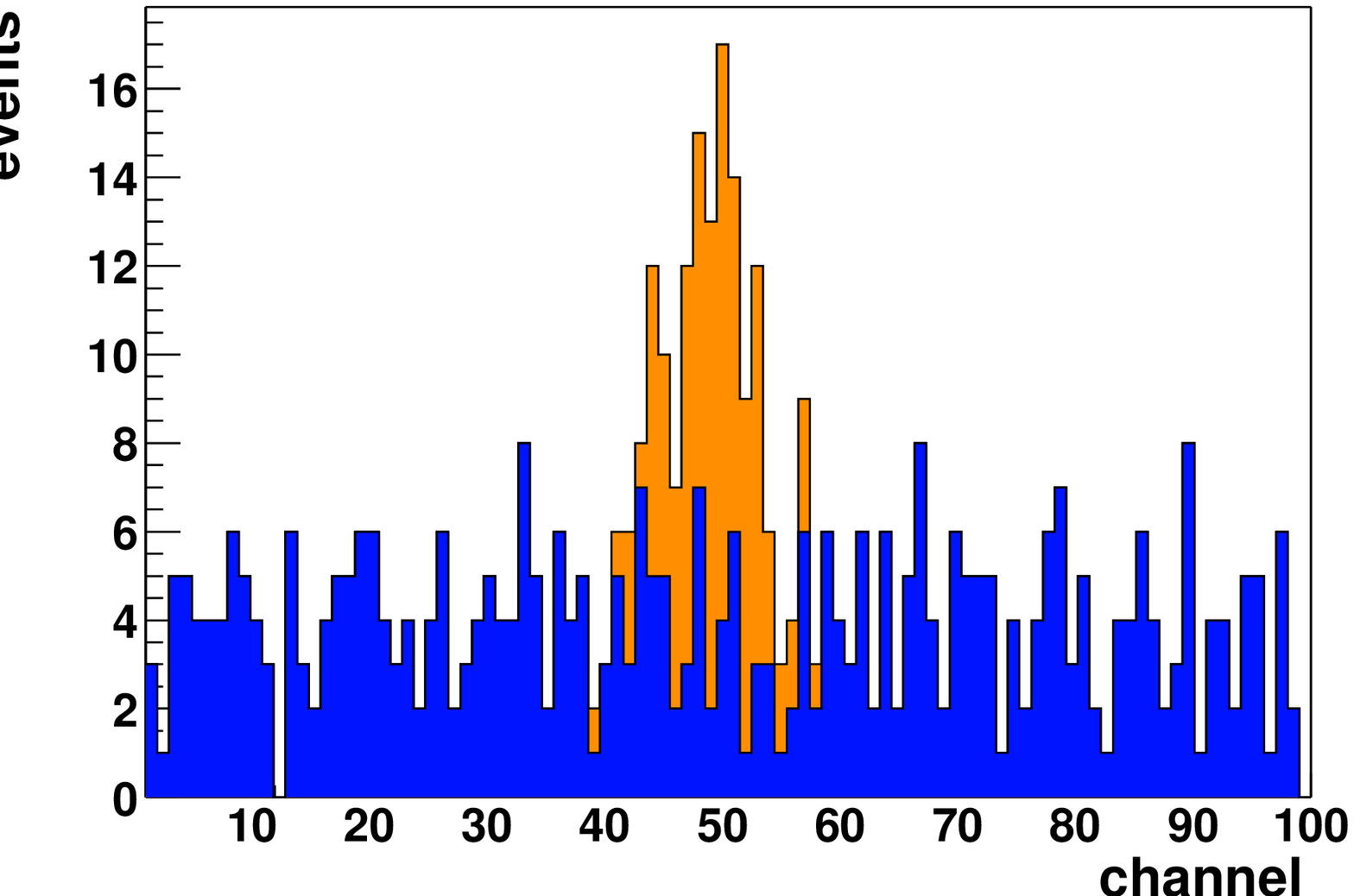}
\end{center}
\caption{\label{picSpecKu}\rm \small 
	Example of a random-generated spectrum with a Poisson distributed
	background with 4.0 events per channel and a Gaussian 
	line centered in channel 50 (line-width corresponds 
	to a standard-deviation of $\sigma=4.0$ channels).
	The left picture shows a spectrum with a line-intensity of 10 events,
	the right spectrum a spectrum with a line-intensity of 100 events.
	The background is shown dark, the events of the line bright.}
\end{figure}


\begin{figure}[th]
\begin{center}
\includegraphics[width=6.7cm]{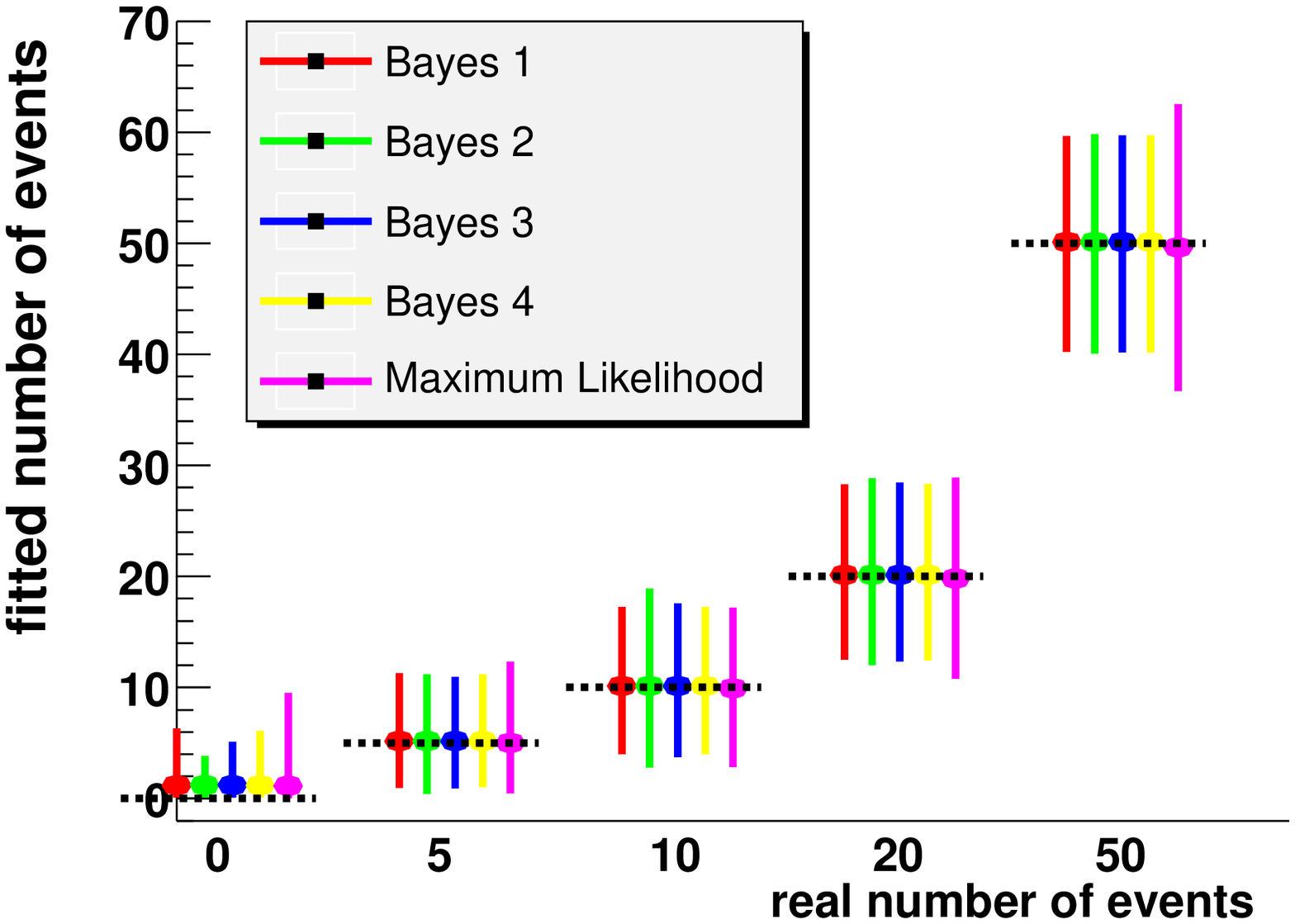}
\includegraphics[width=6.7cm]{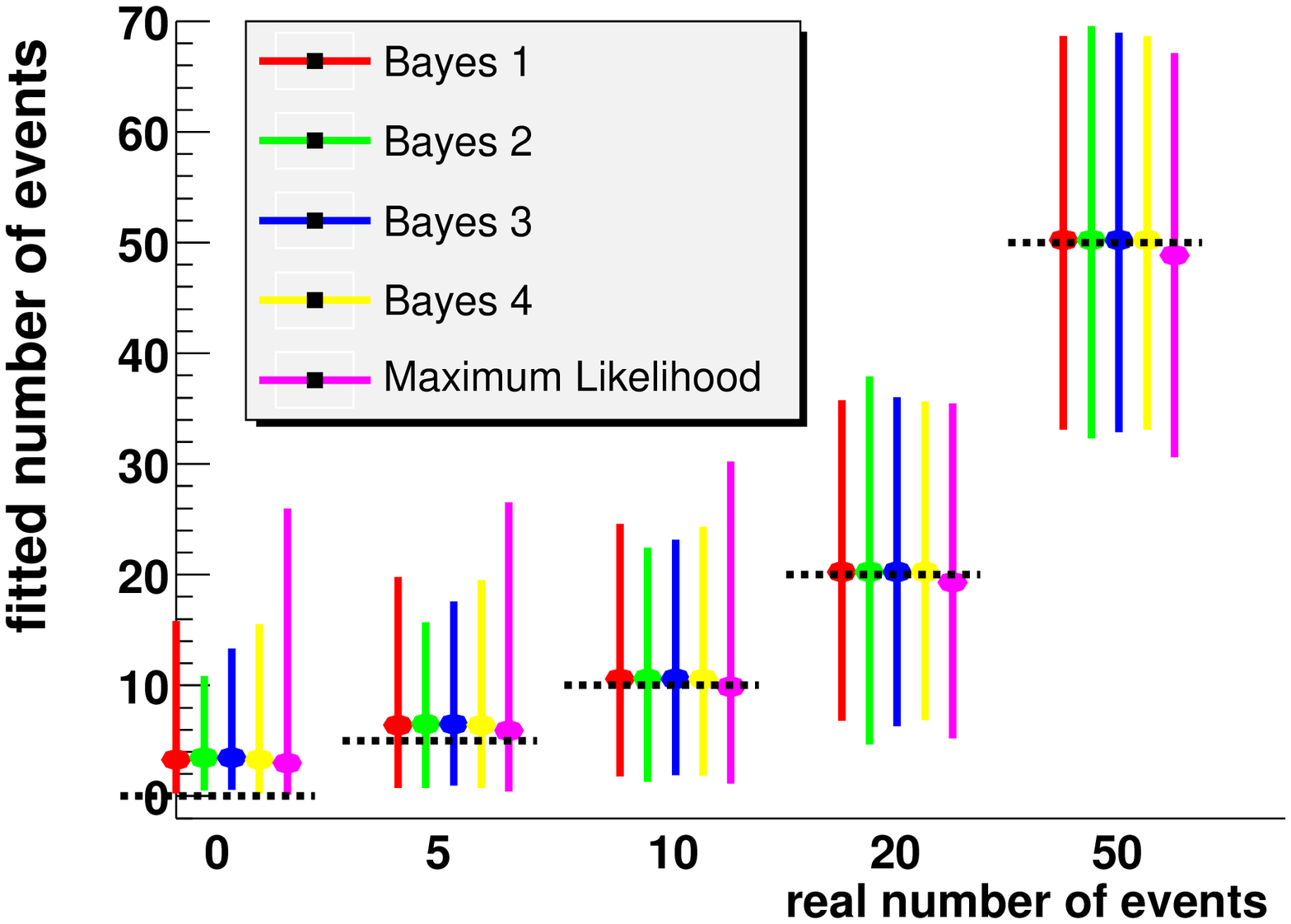}
\end{center}
\caption{\label{picPrior} \rm \small 
	Results of analysis of random-number generated spectra, using Bayes
	and Maximum Likelihood method (the first one with different 
	prior distributions).
	For every number of events in the simulated line, shown 
	on the x-axis, 1000 random generated 
	spectra were evaluated with the five given methods.
	The analysis on the left side was performed with an Poisson
	distributed background of 0.5 events per channel, the background for
	the spectra on the right side was 4.0 events per channel.
	Every vertical line shows the mean value of the calculated best values
	(thick points) with the 1$\sigma$ error area.
	The mean values are in good agreement with the expected values (black
 	horizontal dashed lines).}
\end{figure}


\section{Statistics at Low Count Rates - Peak Search, and Analysis Window}

	At this point it may be useful to demonstrate the potential of the used
	peak search procedure. 
Fig. \ref{picSpecKu} 
	shows a spectrum with Poisson-generated background of
	4 events per channel and a Gaussian line with width (standard
	deviation) of 4 channels centered at channel 50, with intensity of 10
	(left) and 100 (right) events, respectively.
Fig. \ref{picPrior}, 
	right shows the result of the analysis of spectra of
	different line intensity with the Bayes method (here Bayes 
	1-4 correspond to different choice of the prior distribution:
(1) $\mu(\eta)=1$ (flat), (2) $\mu(\eta) = 1/\eta$, 
(3) $\mu(\eta) = 1/\sqrt{\eta}$,
(4) Jeffrey's prior) 
	and the Maximum Likelihood Method.
	For each prior 1000 spectra have been generated 
	with equal background and
	equal line intensity using random number generators available 
	at CERN 
\cite{random}.
	The average values of the best values agree (see Fig. 
\ref{picPrior}) very well
	with the known intensities also for very low count rates (as in Fig.
\ref{picSpecKu}, left).\\

\begin{figure}[htb]
\begin{center}
\includegraphics[width=6.7cm]{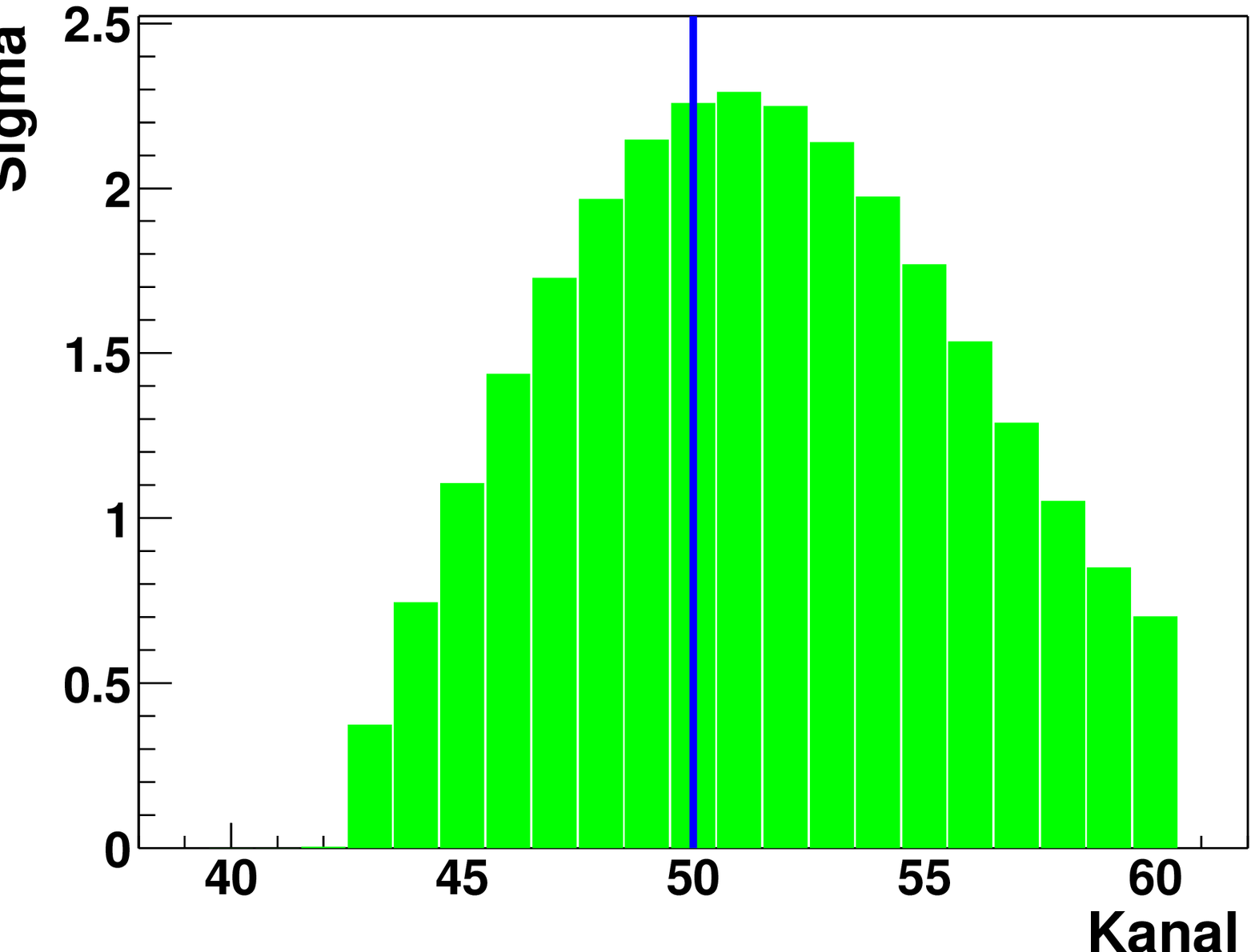}
\includegraphics[width=6.7cm]{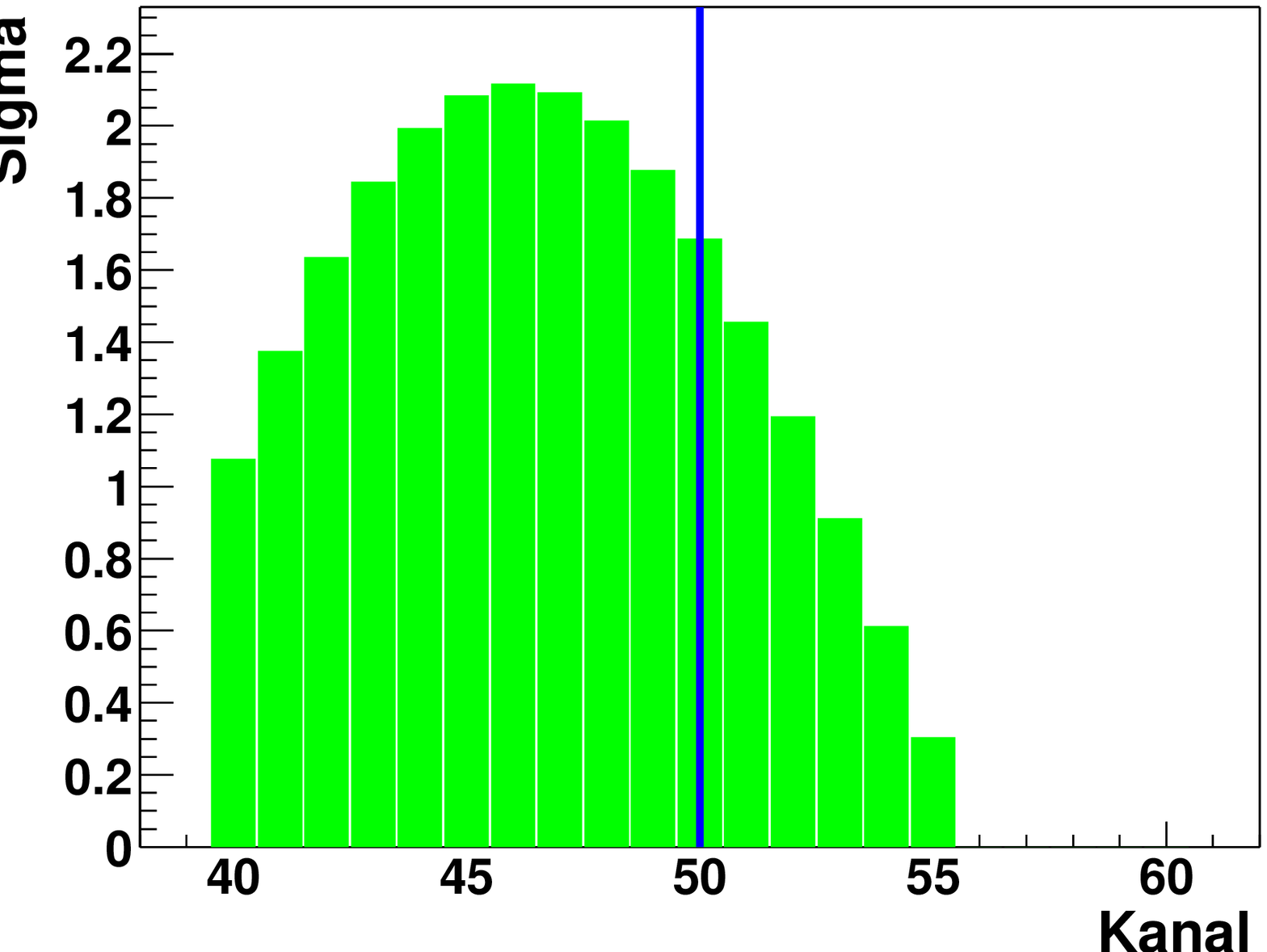}
\caption{\label{picWH1} \rm \small 
	Two spectra with a Poisson-distributed background and a
	Gaussian line with 15 events centered in channel 50 (with a width
	(standard-deviation) of 4.0 channels) created with different 
	random numbers. 
	Shown is the result of the peak-scanning of the spectra.
	In the left picture the maximum of the probability corresponds well
	to the expected value (black line) whereas in the right
	picture a larger deviation is found. 
	When a channel corresponds  to 0.36 keV the deviation in the right
	picture is $\sim$ 1.44 keV.}
\end{center}
\end{figure}


\begin{figure}[h!]
\begin{center}
\includegraphics[width=10cm]{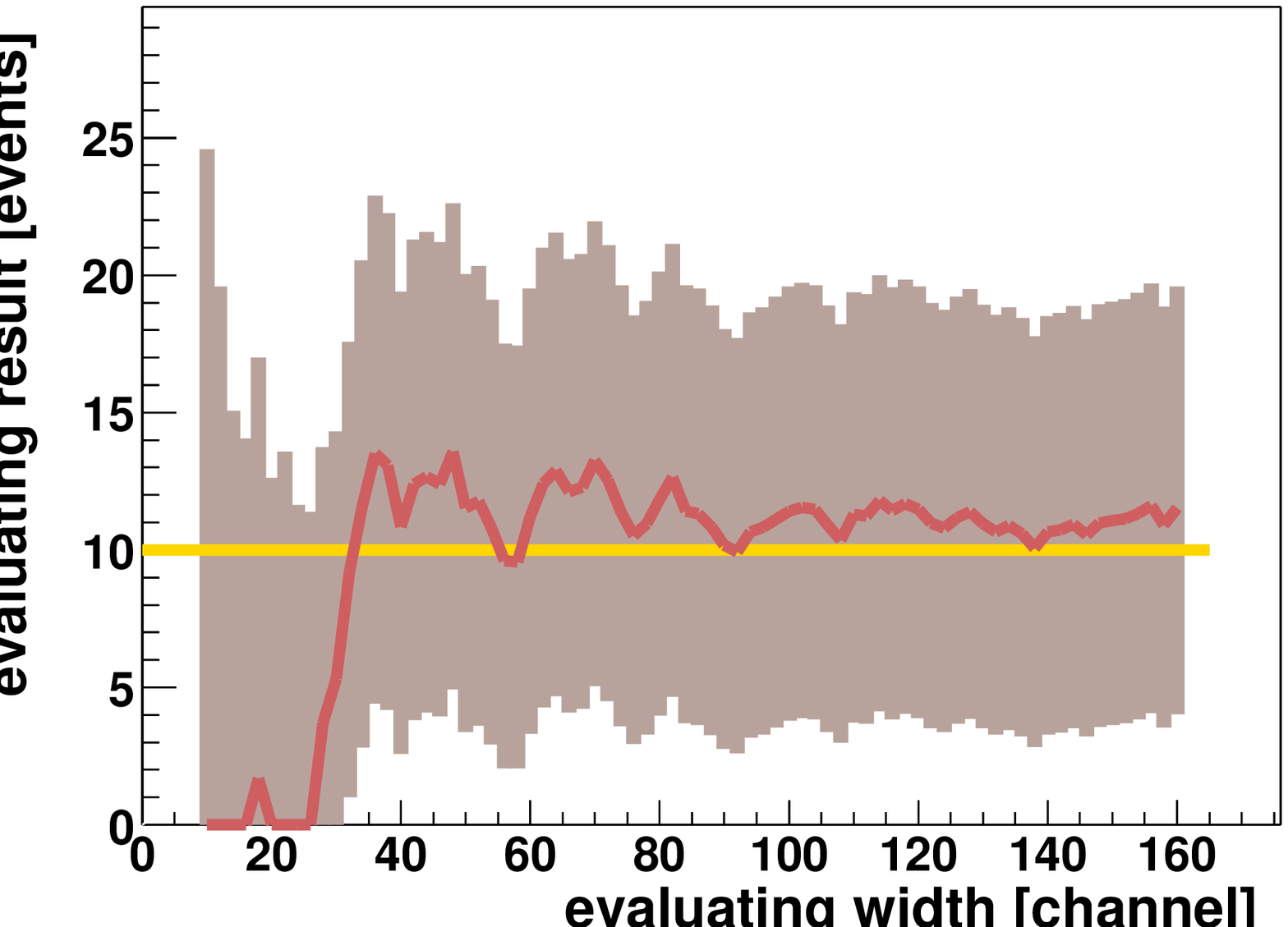}
\end{center}
\caption{\label{picSingleDev} \rm \small 
	Result of an analysis as function of the evaluation width.
	The used spectrum consists of a Poisson distributed background with 4
	events per channel, and a line of 10 events (see fig. \ref{picSpecKu},
	left part). 
	The dark area corresponds to a 68.3\% confidence area with the dark
	line being the best value.
	Below an evaluation width of 35 channels the result becomes unreliable,
	above 35 channels the result is stable.}
\end{figure}


	In Fig. 
\ref{picWH1} 
	we show two simulations of a Gaussian line of 15 events, 
	centered at channel 50, again with width (standard deviation) 
	of 4 channels, on a Poisson-distributed background 
	with 0.5 events/channel. The figure gives an
	indication of the possible degree of deviation of the energy of the
	peak maximum from the transition energy,  on the level of statistics
	collected in experiments like the 
	HEIDELBERG-MOSCOW experiment (here one
	channel corresponds to 0.36 keV). 
	This should be considered when comparing
Figs. \ref{figHM}, \ref{figCaldwell}, \ref{figITEP}, \ref{figIGEX}.
\\


\begin{figure}[th]
\begin{center}
\includegraphics[width=14cm]{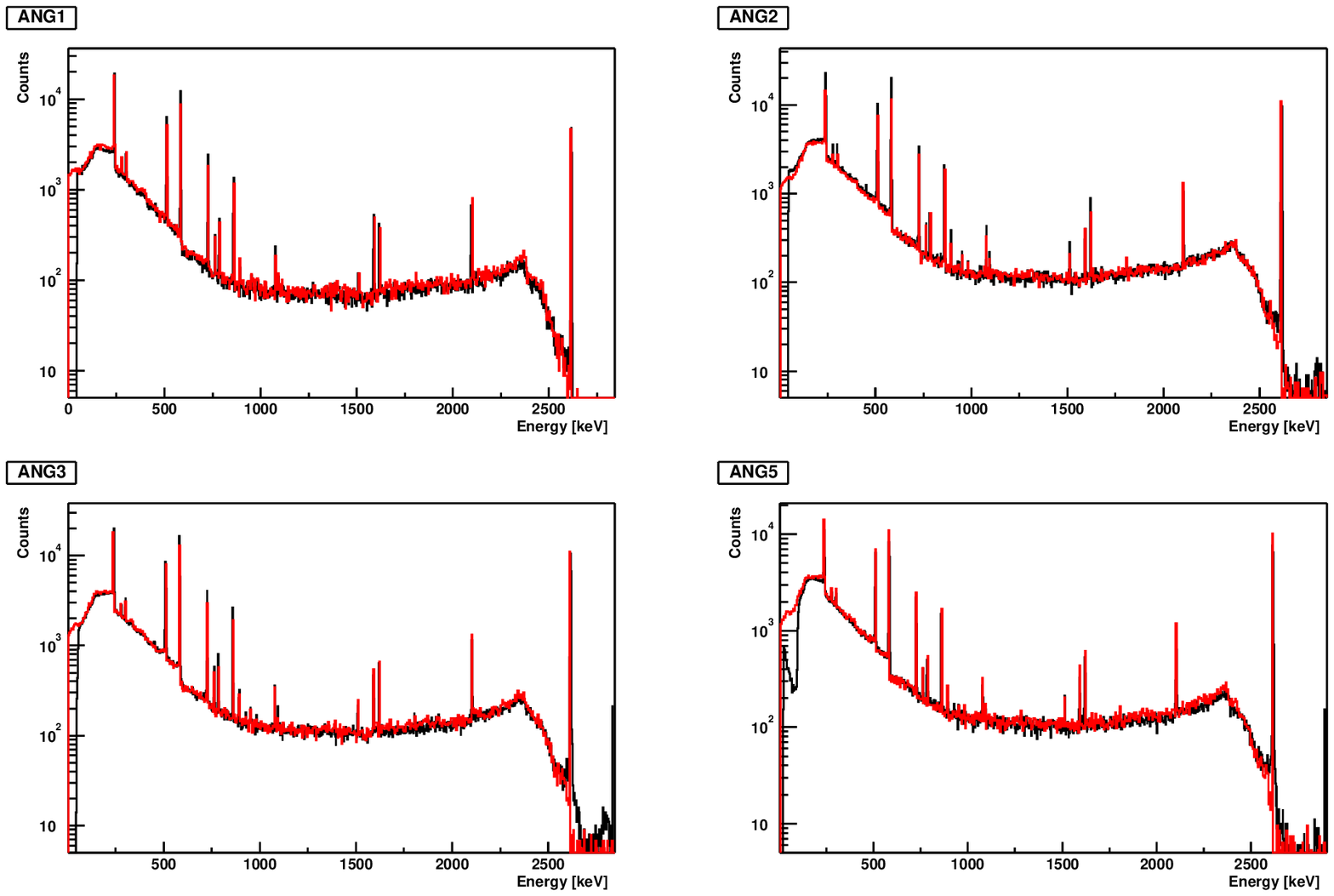}
\caption{\label{picUnderTotal}\rm \small Comparison of the
	measured data (black line, November 1995 to April 2002) and simulated
	spectrum (red line) for the detectors Nrs. 1,2,3 and 5 for a
	$^{232}$Th source spectrum.
	The agreement of simulation and measurement is excellent.}
\end{center}
\end{figure}


\begin{figure}[tb]\vspace*{-0.5cm}
\begin{center}
\includegraphics[width=10.0cm]{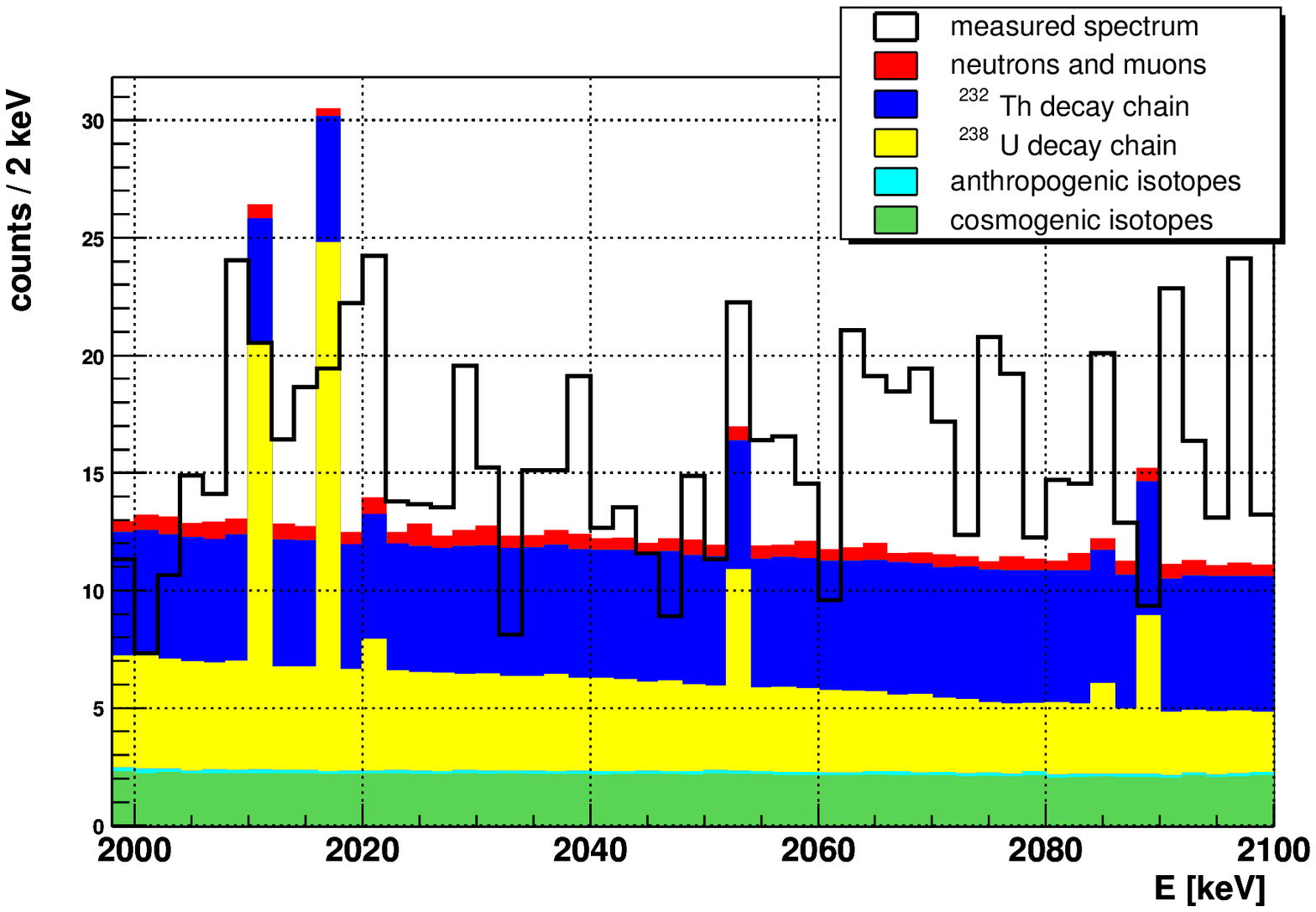}
\caption{\label{picUnder}\rm \small Simulated background of the
	HEIDELBERG-MOSCOW experiment in the energt range from 2000 to 2100 keV 
	with all known background components.
	The black trough-drawn line corresponds to the measured data from
	20.11.1995 to 16.4.2002 (55.57 kg y).}
\end{center}
\end{figure}

	The influence of the choice of the energy range 
	of the analysis around $Q_{\beta\beta}$
	has been thoroughly discussed in \cite{evid2,evid3}. 
	Since erroneous ideas about this
	point are still around, a few further comments may be given here. In
Fig. \ref{picSingleDev} 
	we show the analysis of a simulated spectrum
	consisting of a Gaussian line of width (standard deviation) 
	of 4 channels
	and intensity of 10 counts on a Poisson-distributed background 
	of 4 events per channel (see fig. 
\ref{picSpecKu} left), 
	as function of the width of the range of analysis. 
	It is seen that a reliable result is
	obtained for a range of analysis of not smaller than 35 channels
	(i.e. $\pm$18 channels) - one channel corresponding to 0.36 keV in the
	HEIDELBERG-MOSCOW experiment.
	This is an important result, since it is of course 
	important to keep the
	range of analysis as  \mbox{s m a l l} as possible, to avoid to include
	lines in the vicinity of the weak signal into the background.
	This unavoidably occurs when e.g. proceeding as suggested in Aalseth
	et. al., hep-ex/0202018 and Mod. Phys. Lett. A 17 (2002) 1475,
	Yu.G. Zdesenko et. al.,
	Phys. Lett. B 546 (2002) 206.
	The arguments given in those papers are therefore incorrect. Also
	Kirpichnikov, who states \cite{kirch} that his analysis 
	finds a 2039 keV signal in
	the HEIDELBERG-MOSCOW spectrum on a 4 sigma confidence level 
	(as we also see it, when using the Feldman-Cousins method 
\cite{dietzdiss}), 
	makes this mistake when analyzing the pulse-shape spectrum.\\

\section{Simulation of the Background with \protect \newline GEANT 4}

	Finally the background around $Q_{\beta\beta}$ will 
	be discussed from the side of simulation. 
	A very careful new simulation of the different components of
	radioactive background in the HEIDELBERG-MOSCOW experiment has been
	performed recently by a new Monte Carlo program based on GEANT4 
\cite{cdoerr,ref28}.
	This simulation uses a new event generator for simulation 
	of radioactive decays basing on ENSDF-data and describes 
	the decay of arbitrary radioactive isotopes including alpha, 
	beta and gamma emission as well as
	conversion electrons and X-ray emission. 
	Also included in the simulation is
	the influence of neutrons in the energy range from thermal to high
	energies up to 100 MeV on the measured spectrum. Elastic and inelastic
	reactions, and capture have been taken into account, 
	and the corresponding
	production of radioactive isotopes in the setup. The neutron fluxes and
	energy distributions were taken from published 
	measurements performed in the Gran Sasso. 
	Also simulated was the cosmic muon flux measured in the
	Gran Sasso, on the measured spectrum.
	To give a feeling for the quality of the simulation,
Fig. \ref{picUnderTotal} 
	shows the simulated and the measured spectra for a $^{228}$Th 
	source spectrum for 4 of our five detectors. 
	The agreement is excellent.\\

	The simulation of the background of the experiment reproduces  a l l
	observed lines in the energy range between threshold  (around 100 keV)
	and 2020 keV 
\cite{cdoerr}.  
	Fig. 
\ref{picUnder} 
	shows the simulated background in the range
	2000-2100 keV with all  k n o w n  background components. 
	The black solid line corresponds to the measured data 
	in the period 20.11.1995 - 16.4.2002 (55.57 kg y).\\

	The background around $Q_{\beta\beta}$ is according 
	to the simulations  f l a t, the only expected lines 
	come from $^{214}$Bi (from the $^{238}$U natural decay chain)
	at 2010.89, 2016.7, 2021.6, 2052.94, 2085.1 and 2089.7 keV. 
	Lines from cosmogenically produced $^{56}$Co (at 2034.76 keV 
	and 2041.16 keV), half-life 77.3 days, are not expected since 
	the first 200 days of measurement of
	each detector are not used in the data analysis. 
	Also the potential contribution from
	decays of $^{77}$Ge, $^{66}$Ga, or $^{228}$Ac, 
	should not lead to signals visible in our
	measured spectrum near the signal at $Q_{\beta\beta}$. 
	For details we refer to  
\cite{ref28}.\\

	The structures around 2028 keV, 2066 keV and 2075 keV seen - 
	as also the $^{214}$Bi lines -  in practically all Ge 
	experiments (see above), cannot be identified at present. 
	The 2028 keV line because of its strong occurence 
	in the PSA spectrum in the HEIDELBERG-MOSCOW experiment, 
	may play a special role here (see above).
\\

\section{Conclusion}

	Concluding, additional support has been given for the evidence 
	of a signal for neutrinoless double beta decay, by
	showing consistency of the result - for the signal,  a n d  for the
	background - with other double beta decay experiments using
	non-enriched or enriched Germanium detectors. In particular 
	it has been shown that the lines seen in the vicinity 
	of the signal (including those which at present cannot 
	be attributed) are seen also in the other experiments. 
	This is important for the correct treatment of the background. 
	Furthermore, the sensitivity of the peak identification
	procedures has been demonstrated by extensive statistical simulations.
	It has been further shown by new extensive simulations of the expected
	background by GEANT4, that the background around $Q_{\beta\beta}$ 
	should be flat, and
	that no known gamma line is expected at the energy of $Q_{\beta\beta}$.
	The 2039 keV signal is seen  \mbox{o n l y}  in
	the HEIDELBERG-MOSCOW  experiment, which has a factor of 10, and
\mbox{m u c h} more, statistics than all other double beta experiments.\\

\end{document}